\documentclass[aps,nofootinbib,notitlepage,longbibliography,twocolumn, superscriptaddress]{revtex4-1}
\usepackage{amsmath,amssymb}
\usepackage{slashed}
\usepackage{graphicx}
\usepackage{bm}
\usepackage{float}
\usepackage{dsfont}
\usepackage[T1]{fontenc}
\usepackage{multirow}
\usepackage[utf8]{inputenc}
\usepackage{gauss} 
\usepackage[normalem]{ulem}
\usepackage[top=1.0in,bottom=1.0in,left=1.0in,right=1.0in]{geometry}
\usepackage[colorlinks,linkcolor=blue,citecolor=blue]{hyperref}
\usepackage{subfig}
\usepackage{booktabs}
\usepackage{caption}
\newcommand{\be}{\begin{equation}}
\newcommand{\ee}{\end{equation}}
\newcommand{\bea}{\begin{eqnarray}}
\newcommand{\eea}{\end{eqnarray}}
\newcommand{\ba}{\begin{eqnarray}}
\newcommand{\ea}{\end{eqnarray}}

\def\be{\begin{eqnarray}}
\def\ee{\end{eqnarray}}
\def\bea{\be}
\def\eea{\ee}

\def\roughly#1{\mathrel{\raise.3ex\hbox{$#1$\kern-.75em%
\lower1ex\hbox{$\sim$}}}}

\usepackage{lipsum}

\begin{document}

\title{Quasifragmentation functions in the massive Schwinger model}
\author{Sebastian Grieninger}
\email[]{sebastian.grieninger@stonybrook.edu}
\affiliation{Center for Nuclear Theory, Department of Physics and Astronomy,
Stony Brook University, Stony Brook, New York 11794–3800, USA}
\affiliation{Co-design Center for Quantum Advantage (C2QA), Stony Brook University, Stony Brook, New York 11794–3800, USA}

\author{Ismail Zahed}
\email[]{ismail.zahed@stonybrook.edu}
\affiliation{Center for Nuclear Theory, Department of Physics and Astronomy,
Stony Brook University, Stony Brook, New York 11794–3800, USA}

\date{\today}
\begin{abstract}
We introduce the concept of the quark quasifragmentation function (qFF) using an equal-time and spatially boosted form of  the Collins-Soper fragmentation function where the out-meson fragment is replaced by the current asymptotic condition. We derive the  qFF for a fermion in two-dimensional quantum electrodynamics (QED2) using the Kogut-Susskind Hamiltonian after a mapping onto spin qubits in a spatial lattice with open boundary conditions. This form is suitable for quantum computations. We compute the qFF by exact diagonalization of the spin Hamiltonian. The results are compared to the  qFF following from the Drell-Levy-Yan result for QED2, both at strong and weak coupling, and to two-dimensional quantum chromodynamics in the lowest Fock approximation.
\end{abstract}

\maketitle

\section{Introduction}
Parton distribution and fragmentation functions are central for
the analyses of most high energy data~\cite{Collins:1989gx,Collins:2011zzd}. On the light front, hadrons are composed of frozen partons thanks to time dilatation and asymptotic freedom~\cite{Callan:1969uq,Gross:1973id,Politzer:1973fx}.  
As a result, a hard process in quantum chromodynamics (QCD) can be split into a perturbatively calculable hard block  times non-perturbative matrix elements like parton distribution functions (PDFs) and fragmentation functions (FFs).

The PDFs are valued on the light front and inherently non-perturbative, which makes them inaccessible to standard Euclidean lattice formulations, with the exception of the few lowest moments. This shortcoming is circumvented through the
use of quasi-distributions~\cite{Ji:2013dva}, and their variants~\cite{Radyushkin:2017gjd,Ma:2014jla}.   These proposals have now been pursued by a number of QCD lattice collaborations~\cite{Zhang:2017bzy,Ji:2015qla,Bali:2018spj,Alexandrou:2018eet,Izubuchi:2019lyk,Izubuchi:2018srq}. We have recently shown how these concepts can be extended to quantum computation~\cite{Grieninger:2024cdl}.

The concept of quark fragmentation, originates with the original work by Field and Feynman, who  put forth the quark jet model 
to describe meson production in semi-inclusive processes~\cite{Field:1977fa}. The model is essentially an independent parton cascade
model, where a hard parton  depletes its longitudinal momentum by emitting successive mesons through a chain process. This idea finds a natural realization in the string breaking at the origin of the Lund model~\cite{Andersson:1997xwk}. Jet fragmentation and hadronization are vigorously pursued at current collider facilities both at the Relativistic Heavy Ion Collider and Large Hadron Collider. They be sought at the future Electron-Ion Collider to extract the partonic structure of 
matter, the gluon helicity in nucleons and the mechanism behind the production of diffractive dijets~\cite{Abir:2023fpo,AbdulKhalek:2021gbh}. 

A first principle formulation of the FFs on the light front was suggested by Collins and Soper (CS)~\cite{Collins:1981uw}. They are fully gauge invariant but inherently non-perturbative. Remarkably, the CS FFs are still not accessible  to first principle QCD lattice simulations, due to their inherent light front structure, for a review see~\cite{Metz:2016swz}. A  number of phenomenological and nonperturbative constructions  have exploited crossing and analyticity symmetries, to approximate the FFs from the  PDFs~\cite{Ito:2009zc,Nam:2011hg,Matevosyan:2011vj}, as initially suggested by Drell-Levy-Yan (DLY)~\cite{Drell:1969jm,Drell:1969wd}. 

This work is a follow up on our original work for the quasi-PDF (qPDF) using exact diagonalization of the spin Hamiltonian~\cite{Grieninger:2024cdl}. We will show that a first principle and non-perturbative analysis of the CS light front FFs can be  obtained through  suitably boosted quasi-FFs (qFFs) defined at spatial separation, where the out-going meson fragment is replaced by the current asymptotically.
qFFs are readily implemented on
a quantum computer using massive two-dimensional QED (QED2), the Kogut-Susskind Hamiltonian~\cite{Kogut:1974ag}  on a spatial lattice with open boundary conditions.  The fermion fields are eliminated by a standard Jordan-Wigner transformation~\cite{Jordan:1928wi}. Note that  partonic physics on quantum computers was also recently studied in~\cite{Lamm:2019uyc,Echevarria:2020wct,Kreshchuk:2020aiq,Perez-Salinas:2020nem,Li:2021kcs,Qian:2021jxp,Li:2022lyt,Zache:2023cfj}.

This paper contains a number of new results: 1) an equal-time formulation of 
the FF for increasing boosts or qFF, that
asymptotes the CS FF in the light-like limit; 2) explicit derivation of the DLY relation for 2D gauge theories in the light-like limit; 3) a qubit
form of the qFF in QED2 for increasing boosts; 4) a numerical evaluation of 
the spatial correlator for the qFF in 
QED2 for increasing time and boosts and its comparison to the luminal CS FF. These new theoretical steps and checks are essential in assessing the
FFs in a gauge theory prior to their implementation on quantum hardware.

The outline of the paper is as follows: in section~\ref{SECII}, we briefly outline
the essentials of massive QED2 for this work. We introduce the concept of the quark qFF by recasting the CS light-front FF into a spatial correlator with properly boosted out-meson states. In section~\ref{SECIII}, we show how the approximate  DLY relation can be used to extract the   qFFs from the qPDFs for QED2. 
In section~\ref{SECIV}, we proceed to discretize the qFF in QED2 using the Kogut-Susskind Hamiltonian on a spatial lattice with open boundary conditions. The fermion fields are eliminated with the help of a Jordan-Wigner transformation, and the ensuing qFF is shown to asymptote the CS FF for increasing boosts. Note that this formulation of the qFF can be used to perform quantum simulations on quantum hardware. Our conclusions are in section~\ref{SECV}. In the appendixes, we review a numerical algorithm for QED2 and
discuss the DLY result for QCD2.

\section{Massive QED2}
\label{SECII} 
Two-dimensional QED (QED2)  also known as the Schwinger model~\cite{Schwinger:1962tp}, exhibits a variety of non-perturbative phenomena familiar from 4-dimensional 
gauge theories. The extensive interest in QED2 stems from the fact that it bears much in common with two-dimensional QCD, with Coulomb law confining in two dimensions. As a result, the QED2 spectrum involves only chargeless excitations. Remarkably, the vacuum state is characterized by a non-trivial chiral condensate and topologically active tunneling configurations. The massive Schwinger model is not exactly solvable and has recently received a lot of interest as a playground for quantum computations on classical and quantum hardware, see for example \cite{Klco:2018kyo,Farrell:2023fgd,Farrell:2024fit,Zache:2018cqq,Rigobello:2021fxw,Kharzeev:2020kgc,Florio:2023dke,Lin:2024eiz,Lee:2023urk,deJong:2021wsd,Belyansky:2023rgh,Barata:2023jgd,Lee:2023urk,Florio:2023mzk} and references therein. Moreover, entanglement in Schwinger pair production was investigated in~\cite{Berges:2017zws,Berges:2017hne,Florio:2021xvj,Grieninger:2023ehb,Grieninger:2023pyb}.

QED2 with massive fermions is described by~\cite{Schwinger:1962tp,Coleman:1976uz}
\bea
\label{A1}
S=\int d^2x\,\bigg(\frac 14F^2_{\mu\nu} +\frac{\theta \tilde F}{2\pi}+\overline \psi (i\slashed{D}-m)\psi\bigg)
\eea
with $\slashed{D}=\slashed{\partial}-ig\slashed{A}$. The bare fermion mass is $m$ and the coupling $g$ has mass dimension. At weak coupling with $m/g>1$, the spectrum is composed of heavy mesons, with doubly degenerate C-even and C-odd vacuua at $\theta=\pi$.   At strong coupling  with $m/g<1$, the spectrum is  composed of light mesons and baryons with a C-even vacuum independent of $\theta$. Moreover, as we recently discussed in \cite{Grieninger:2023ufa}, in the strong coupling limit the squared rest mass may be expressed as
\bea
m_\eta^2=m_S^2+m_\pi^2=\frac {g^2}\pi-\frac{m\langle \overline{\psi}\psi\rangle_0}{f_\eta^2}
\eea 
with the anomalous contribution to the squared mass $m_S^2=g^2/\pi$
and the $\eta$ decay constant $f_\eta=1/\sqrt{4\pi}$. 
The vacuum chiral condensate is finite in the chiral limit with
$\langle\overline\psi \psi\rangle_0=-\frac{e^{\gamma_E}}{2\pi}m_S$
~\cite{Sachs:1991en,Steele:1994gf}.
At weak coupling, the squared mass is expected to asymptote $(2m)^2$.

\subsection{Quark FF}
On the light front,  a gauge-invariant definition of the QCD  quark fragmentation $Q\rightarrow Q+H$ was
given by Collins and Soper in~\cite{Collins:1981uw}.
When reduced to QED2 it reads
\begin{widetext}
\begin{equation}
\label{LFX}
  d_{q}^\eta(z,1)=\frac 1z\,\int\frac{dz^-}{4\pi}e^{-iz^{-1}P^+z^-}
\,{\rm Tr}\bigg(\gamma^+\gamma^5\langle 0|
\psi(0^-) [0^-,\infty^-]^\dagger a_{\rm out}^\dagger(P^+)a_{\rm out}(P^+)
[\infty^-,z^-]\overline\psi(z^-)|0\rangle
\bigg),
\end{equation}
\end{widetext}
with a quark of longitudinal momentum $k^+=P^+/z$ fragmenting into an on-shell 
$H=\eta'$ with longitudinal momentum $zk^+=P^+$. The out-$\eta$ mode annihilation operator is $a_{\rm out}$.
The holonomy  along the light cone
\bea
[x^-, y^-]={\bf P}\bigg({\rm exp}\bigg(-ig\int_{x^-}^{y^-}dz^-A^+(z^-)\bigg)\bigg)
\eea
 can be omitted in the light-cone gauge. 
Eq. (\ref{LFX}) can be regarded as the number of $\eta$s of momentum $P^+$ in a fast moving quark (dressed by a gauge holonomy) of momentum $P^+/z$. Our conventions for the gamma
matrices are $\gamma^0=\sigma^3$ and $\gamma^1=i\sigma^2$, with $\gamma^+=\gamma^0+\gamma^1$ and  $\overline\psi\gamma^+\gamma^5\psi=\overline\psi\gamma^+\psi$. 
 
 Using the ``good'' component of the fermionic field in light front quantization,
and setting the gauge links to 1,  (\ref{LFX}) can be recast as
 \bea
 d_\eta^q(z)=\frac{dP^+}{dz}\,\frac{\langle k^+|a_{\rm out}^\dagger(P^+)a_{\rm out}(P^+)|k^+\rangle}{\langle k^+|k^+\rangle}
 \eea
 which is the light cone momentum distribution of the meson $\eta$ in the 
 quark $q$ (dressed by the holonomy). It satisfies the momentum sum rule
 \bea
 \label{SUM}
  \int_0^1dz\, zd_\eta^q(z)=1
  \eea
provided that the quark state is an eigenstate of the {\bf mesonic} momentum operator, 
\bea
\bigg(\mathbb P=\int dP^+\,P^+\,a_{\rm out}^\dagger(P^+)a_{\rm out}(P^+)\bigg)|k^+\rangle=k^+|k^+\rangle\nonumber\\
\eea
In the absence of a conserved charge in massive QED2, 
\bea
  \int_0^1dz\,d_\eta^q(z)=
  \frac{\langle k^+|\mathbb Q=\int dP^+\,a_{\rm out}^\dagger(P^+)a_{\rm out}(P^+)|k^+\rangle}{\langle k^+|k^+\rangle}\nonumber\\
\eea
is not normalized to 1, as the number of out-going mesons is not fixed. The generalization to multiflavor QED2 would lead to conserved flavor charges.

\begin{widetext}
  \subsection{Quark qFF}
 With this in mind, we introduce the  qFF  for QED2
\bea
\label{DPRIMITIVE}
  d_{q}^\eta(z,v)=\frac 1z\,\int\frac{dZ}{4\pi}e^{-i\frac 1zP(v)Z}
\,{\rm Tr}\bigg(\gamma^+\gamma^5\langle 0|
\psi(0) [0,\infty]^\dagger a_{\rm out}^\dagger(P(v)) a_{\rm out}(P(v))
[\infty, Z]\overline\psi(Z)|0\rangle
\bigg) 
\eea
or the equivalent but spatially symmetric form
\bea
\label{DPRIMITIVESYM}
  d_{q}^\eta(z,v)=\frac 1z\,\int\frac{dZ}{4\pi}e^{-i(\frac 2{z}-1)P(v)Z}
\,{\rm Tr}\bigg(\gamma^+\gamma^5\langle 0|
\psi(-Z) [-Z,\infty]^\dagger a_{\rm out}^\dagger(P(v)) a_{\rm out}(P(v))
[\infty, Z]\overline\psi(Z)|0\rangle
\bigg) 
\eea
\end{widetext}
where $P(v)=\gamma(v) m_\eta v$ is the momentum fraction carried by  the emitted $\eta$ from a mother  quark jet with momentum $P(v)/z$.
Eq. (\ref{DPRIMITIVE}) is seen to reduce to the light front FF (\ref{LFX}) 
in the luminal limit $v\rightarrow 1$. Here the holonomies are running along the spatial direction. In two dimensions, both gauge degrees of freedom $A^{0,1}$ can be eliminated 
by our gauge choice $A^0=0$, and using Gauss law which will be carried out below.

The  out-$\eta$ field canonical decomposition in normal modes is given by
\bea
\!\eta_{\rm out}(x)\!=\!\!
\int\! \frac{dk}{4\pi k^0}
(a_{\rm out}(k)e^{-ik\cdot x}\!+\!a_{\rm out}^\dagger(k)e^{+ik\cdot x})
\eea
with the on-shell energy $k^0=E_k=\sqrt{k^2+m_\eta^2}$.
The asymptotic current condition
\bea
\overline\psi\gamma_\mu\gamma_5\psi(x)\rightarrow
F\partial_\mu\eta_{\rm out} (x)\qquad
x^0\rightarrow +\infty\nonumber\\
\eea
with $F=\sqrt 2 f$, defines
\bea
\frac{Fiq_\mu}{2q^0}\,a_{\rm out}^\dagger(q)=
\int dx e^{-iE_qt+iqx}\overline\psi\gamma_\mu\gamma_5\psi(t,x)|_{t\rightarrow +\infty}\nonumber\\
\eea
or equivalently
\bea
a_{\rm out}^\dagger(q)=
\frac{-2i}{F}\,e^{-iE_kt}
\psi^\dagger\gamma_5\psi (t,q)|_{t\rightarrow +\infty}.
\eea
The asymptotic time limit implements the Lehmann-Symanzik-Zimmermann (LSZ) reduction on the source field.
If we denote the (normal ordered) Hamiltonian operator in  temporal gauge by $\mathbb H$, then we find
\bea
&&a_{\rm out}^\dagger(P) a_{\rm out}(P)=\nonumber\\
&&\frac{4}{F^2}e^{i\mathbb H t}
|\psi^\dagger\gamma_5\psi(0,P(v))|^2
e^{-i\mathbb H t}|_{t\rightarrow +\infty}.
\nonumber\\
\eea
The symmetric qFF (\ref{DPRIMITIVESYM}) can be recast in terms of the spatial qFF correlator
\bea
\label{DPRIMITIVEX}
  d_{q}^\eta(z,v)=\frac 1z\,\int\frac{dZ}{4\pi}e^{-i(\frac 2z-1)P(v)Z}\,\mathbb C(Z,v,\infty)
  \eea
with 
\begin{widetext}
\bea
\label{CVT}
\mathbb C(Z,v,t)\!=\!\frac 4{F^2}{\rm Tr}\bigg(\!\gamma^+\gamma^5
\langle 0|\psi(0,-Z) [-Z,\infty]^\dagger e^{i\mathbb H t} e^{i\chi(v)\mathbb K}
|\psi^\dagger \gamma_5\psi(0,0)|^2
 e^{-i\chi(v)\mathbb K} e^{-i\mathbb Ht}[\infty,Z]\overline\psi(0,Z)|0\rangle\!\bigg).
\eea
\end{widetext}
In QED2 in the temporal gauge $A^0=0$, the Hamiltonian and boost operators associated to (\ref{A1}) are
\bea
\label{HBX}
\mathbb H&=&\int dx\,
\bigg(\frac 12 E^2+\psi^\dagger (i\alpha D_x+m\gamma^0)\psi\bigg),\nonumber\\
\mathbb K&=&\int dx\,x\,
\bigg(\frac 12 E^2+\psi^\dagger (i\alpha D_x+m\gamma^0)\psi\bigg)
\eea
and satisfy the Poincare algebra
\bea
\label{KHP}
[\mathbb K, \mathbb H]&=&i\mathbb P,\\
\left[\mathbb K, \mathbb P\right]&=&i\mathbb H,
\eea
with the momentum operator $\mathbb P$. The rapidity is defined as
\bea
\chi(v)=\frac 12 {\rm ln}\bigg(\frac {1+v}{1-v}\bigg).
\eea
Note that under the combined boost and time evolution, the equal-time fermion field
is now lying on the light cone, e.g.
\bea
&& e^{-i\chi(v)\mathbb K} e^{-i\mathbb Ht}\overline\psi(0,\pm Z)e^{i\mathbb Ht}
 e^{i\chi(v)\mathbb K} \nonumber\\
 &&=\overline\psi(-\gamma(v)(t\pm vZ),\gamma(v)(vt\pm Z))S^{-1}[v]\nonumber\\
 \eea
with $S[v]=e^{\frac 12 \chi(v)\gamma^5}$, for which (\ref{CVT}) in the temporal gauge reads 
\begin{widetext}
\bea
\label{CVTXX}
\mathbb C(Z,v,t)=\frac 4{F^2}{\rm Tr}\bigg(\gamma^+\gamma^5
\langle 0|\psi(-\gamma(v)(t- vZ),\gamma(v)(-Z+vt))
|\psi^\dagger \gamma_5\psi(0,0)|^2
\overline\psi(-\gamma(v)(t+ vZ),\gamma(v)(vt+ Z))|0\rangle\bigg).\nonumber\\
\eea
\end{widetext}

\section{Drell-Levy-Yan in 2D}
\label{SECIII}
Crossing symmetry and charge conjugation allow for an estimate of the CS FF in terms of the parton distribution functions using the DLY relation~\cite{Drell:1969jm}. This observation carries to the primitive qFF with the DLY relation 
\bea
\label{DLY1}
d_{DLY}(z,v)=\frac{z^{d-3}}{de}\,p_\eta \bigg(\frac 1z, v\bigg)
\eea
where $de$ is a pertinent  degeneracy factor which is 1 in QED2
and $N_c$ in QCD2.  
For QED2, the PDF is briefly reviewed in Appendix~\ref{DAPDF} and the qPDF in~\cite{Grieninger:2024cdl}. 
Inserting (\ref{DLY1}) into (\ref{DPRIMITIVEX}) and inverting yield the DLY spatial correlator ($\bar z=1-z$)
\bea
\label{DPRIMITIVEXX}
\mathbb C_{DLY}(Z,v,\infty)=4P\!\int\! dz\,e^{i(z-\bar z)P(v)Z}\,\big|\varphi_\eta\big(z, v\big)\big|^2\!.\nonumber\\
  \eea

\begin{figure}
   \centering
\includegraphics[height=5cm,width=0.99\linewidth]{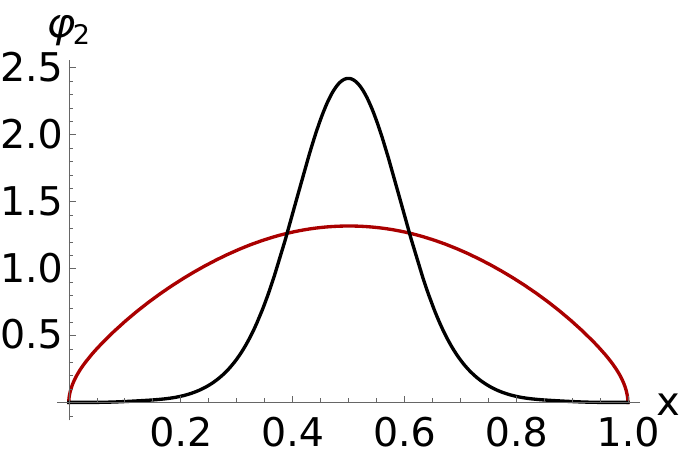}
    \caption{Solution to \eqref{TH1X} for $\beta=\sqrt{3}/\pi$ (red) and $\beta=10\sqrt{3}/\pi  $ (black) using 13 Jacobi polynomials.}
   \label{qfunct}
\end{figure}

The DYL relation can be readily applied to QED2 using the PDFs or qPDFs in~\cite{Grieninger:2024cdl}. Since the qPDFs were obtained using an expansion in 
Jacobi polynomials of increasing powers,  the substitution $x\rightarrow 1/z$
is increasingly singular. To circumvent this, we can use the exact equation (\ref{TH1X})  with the normalized solution $\int dx\,\varphi(x)=f$,
to rewrite (\ref{DLY1}) as
\begin{widetext}
\bea
\label{D1FINITE}
d_{DLY}(z,1)=\frac{\bar z^2}
{z(\bar z\mu^2+z^2\bar\alpha)^2}
\bigg(f-\int_0^1dx \frac{\varphi(x)}{(x-1/z)^2}\bigg)^2
\eea
\end{widetext}
with $\mu^2=M^2/m_S^2$ and $1+\bar\alpha=\alpha=m^2/m_S^2$. We have dropped the principal part prescription since $0\leq z\leq 1$ with $\varphi(x)$ vanishing at the end points. A sample of the light front wave functions solution to (\ref{TH1X}) is shown in Fig.~\ref{qfunct} for strong (solid red line) and weak coupling (solid black line).

\subsection{Light masses}
 In the massless limit, (\ref{D1FINITE}) simplifies to
\bea
\label{MASSLESS}
d_{DLY}(z,1)\rightarrow \frac 1z
\eea
following the substitutions
$\varphi(x)\rightarrow \theta(x\bar x)$,
$f\rightarrow 1$, $\mu^2\rightarrow 1$ and $\bar\alpha\rightarrow -1$.
The apparent singularity in the denominator of (\ref{D1FINITE}) at $\bar z=z^2$,  is canceled by the numerator. We also expect this cancellation to happen in the massive case provided that the exact wave functions  are used.
 For small masses $m/m_S<1$, (\ref{D1FINITE}) scales as $d_\eta(z\sim 0, 1)\sim 1/z$ 
 and vanishes  as $d_\eta(z\sim 1, 1)\sim \bar z^{2\beta}$ since 
 \bea
 &&\bigg(f-\int_0^1dx \frac{\varphi(x)}{(x-1/z)^2}\bigg)_{z\rightarrow 1}^2  \nonumber\\
 &&\approx\bar z^{2(\beta -1)}\bigg(\int_0^\infty ds\frac {s^\beta}{(1+s)^2}\bigg)^2
 \eea
 following the variable shift  $\bar x=s\bar z$ in the integral.
 \begin{figure}
    \centering
\includegraphics[height=5cm,width=0.99\linewidth]{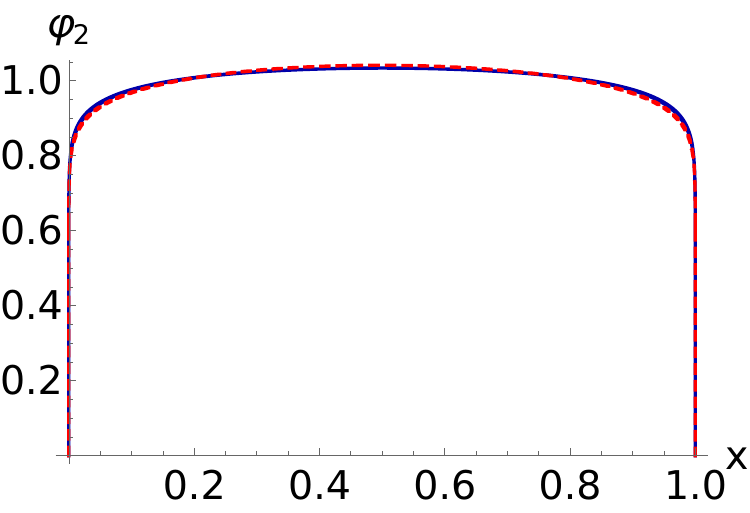}
    \caption{Lowest state wave function in QED2 using the 0th Jacobi polynomial (solid red line) and
    a basis with 13 Jacobi polynomials (solid blue line).}
   \label{COMP}
\end{figure}

 A useful approximation for light quark masses is to use the leading
0th Jacobi polynomial in (\ref{FBARN}) (solid red line) which compares well with the
solution obtained using 13 Jacobi polynomials (solid blue line) for the ground state as shown in Fig.~\ref{COMP}. 
When inserted in the DLY result, the 
approximated form is
\bea\label{eq:deta}
d_{DLY}(z,1)\rightarrow 
\frac{\Gamma(2-2\beta)}{\Gamma(1-4\beta)\Gamma(1+2\beta)}\,\frac{\bar{z}^{2\beta}}{z^{1+4\beta}}
\eea
with the normalization adjusted to satisfy the momentum sum rule~\eqref{SUM}. The solution~\eqref{eq:deta} also shows the expected end point behavior in the strong coupling regime, with $\beta$ satisfying 
$$\beta=\frac{\sqrt 3}\pi\frac m{m_S}< \beta_c=\frac{\sqrt 3}{\pi\sqrt \pi}.$$

\begin{figure}
    \centering
\includegraphics[height=5cm,width=0.99\linewidth]{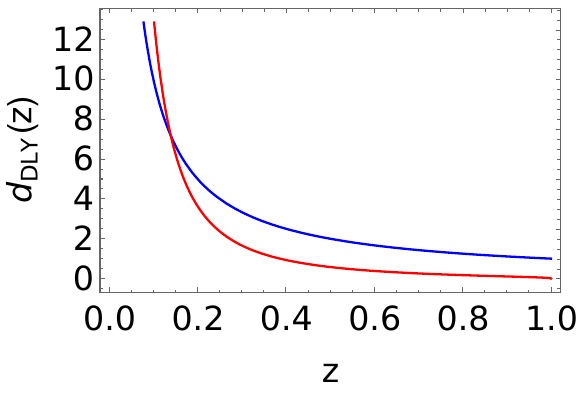}
    \caption{DLY fragmentation function for light quarks in the strong coupling regime: 
    $\beta=0$ (solid blue) and $\beta=0.2$ (solid red) as in (\ref{eq:deta}).}
   \label{DLYLIGHT}
\end{figure}

In Fig.~\ref{DLYLIGHT} we show the behavior of (\ref{eq:deta}) for $\beta=0$ (solid blue)
and $\beta=0.2$ (solid red) in the strong coupling regime. The fragmentation function vanishes in the forward direction $z\rightarrow 1$, and diverges as $z\rightarrow$ 0. The divergence in the $m/m_S\rightarrow 0$ limit is in agreement with the exact bosonization  description
in QED2~\cite{Casher:1974vf}.  In this dual limit, the FF follows from coherent emission from a classical field sourced by light-like currents using bosonization.

\subsection{Heavy masses}
\begin{figure}
    \centering
    \includegraphics[width=0.8\linewidth]{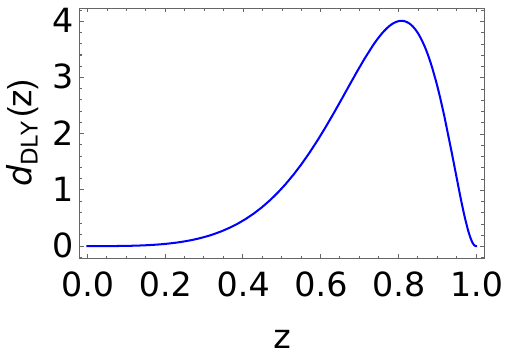}
    \caption{DLY fragmentation function for  heavy  quarks in QED2 as in (\ref{HEAVYQED2}). }
    \label{DLYQED2}
\end{figure}

In the heavy quark limit
$m/m_S\gg 1$,
the anomaly drops out which amounts of
dropping the $f$-contribution in the bracket in (\ref{D1FINITE}). In this limit, we have  $\varphi(x)\rightarrow f_H\delta(x-1/2)$, 
$\mu^2\rightarrow 4\alpha$ and $\bar\alpha\rightarrow \alpha$, with the result
\bea
\label{HEAVYQED2}
d_{DLY}(z,1)\rightarrow 
\frac{16f_H^2}{\alpha^2}\frac{z^3\bar z^2}{(z-2)^8}.
\eea
The constant $f_H$ is fixed by the momentum sum rule
\bea 
\int_0^1 dz\,zd_{DLY}(z,1)= \frac{2f_H^2}{105\,\alpha^2}\rightarrow 1.
\eea
In Fig.~\ref{DLYQED2}, we show the result (\ref{HEAVYQED2}) with the vanishing of the FF at both end points. For heavy fermions, the  FF is peaked in the forward (jet) direction, with a strong suppression as $z\rightarrow 0$.

To address the fragmentation for heavy but finite mass in the 2-particle Fock space,  a useful approximation is to map the light-front
equation for the wave functions onto a standard  Schr\"odinger equation. This
is achieved by noting that (\ref{TH1X}) follows from the light-front quantization
of a two dimensional string with massive end points much like in QCD2~\cite{Bardeen:1975gx,Bars:1976nk}, modulo the U(1) anomaly which drops out in the heavy mass limit. With this in mind, the equal-time quantization for large masses yields~\cite{Bardeen:1975gx,Bars:1976nk,Shuryak:2021hng}
\bea
-\frac 1{m}\varphi_n''(r)+2\pi m_S^2|r|\varphi_n(r)=(E_n-2m)\varphi_n(r)
\eea
with $r$ being the relative separation between the heavy pair. The solutions are Airy functions ($r\geq 1 $)
\bea
\varphi^+_n(r)={\rm Ai}\bigg((2\pi m_S^2 m)^{\frac 13}(r-(E_n-2m)/(2\pi m_S^2))\bigg)
\nonumber\\
\eea
and $\varphi_n^-(r)=(-1)^n\varphi^+_n(-r)$ for $r<0$, with eigen-energies
\bea
E_n=2m-\bigg(\frac{(2\pi m_S^2)^2}m\bigg)^{\frac 13} r_n
\eea
where $r_n$ is the $n$th zero of the Airy function or its derivative. The relationship to the distribution amplitudes (DAs) follows by Fourier inverse
\bea
\varphi_n(\xi)=\int_{-\infty}^{+\infty}\frac{ds}{2\pi}\,e^{-is\xi}\,(\theta(s)\varphi^+_n(s)+\theta(-s)\varphi^-_n(s))\nonumber\\
\eea
after the rescaling $s=m_Sr$, with the symmetric parton fraction $\xi=2x-1$.

\begin{figure*}
\centering
\subfloat[\label{RCZ}]{%
\includegraphics[height=5cm,width=7.5cm]{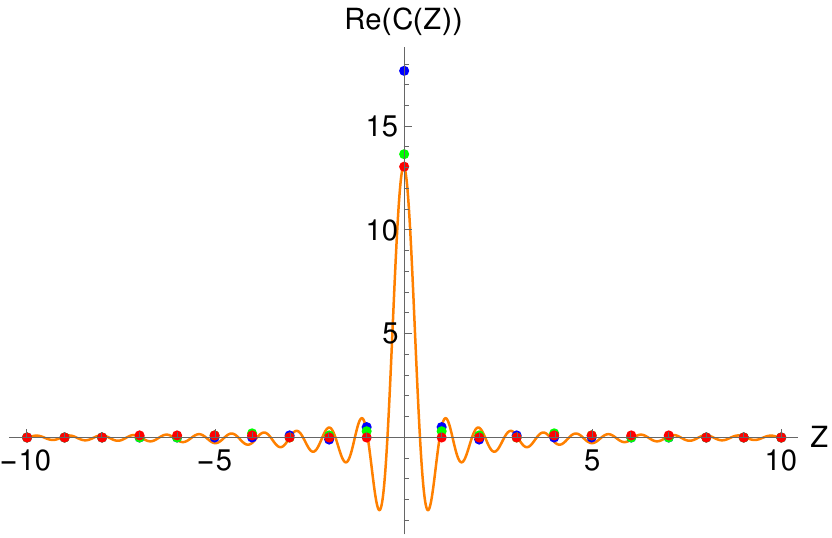}  
}\hfill
\subfloat[\label{ICZ}]{%
\includegraphics[height=5cm,width=7.5cm]{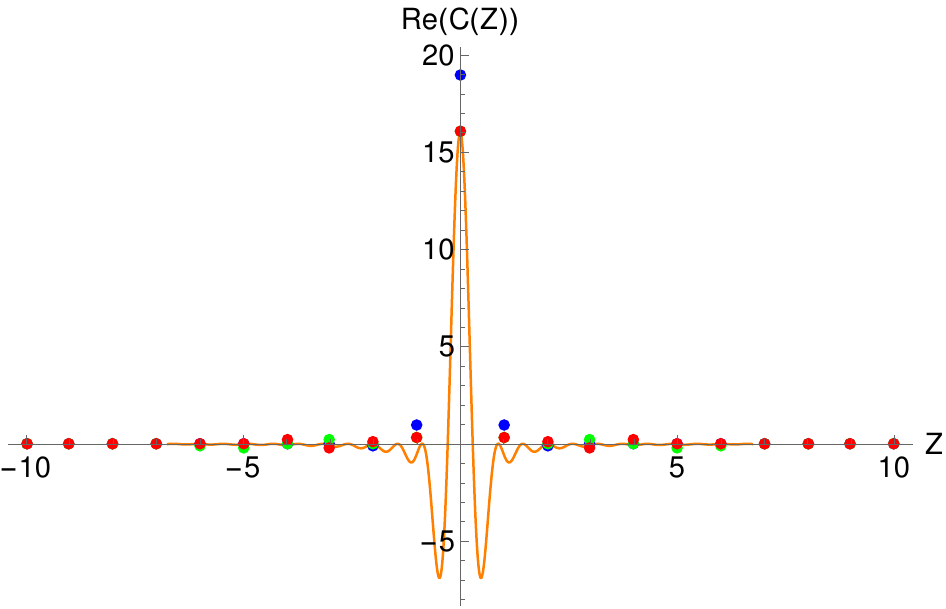}
}\hfill
\subfloat[\label{RCZH}]{%
\includegraphics[height=5cm,width=7.5cm]{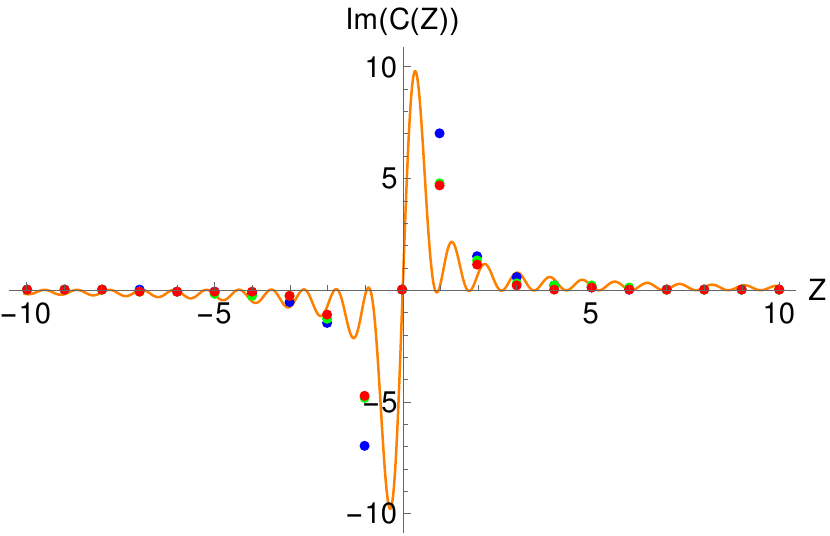}  
}\hfill
\subfloat[\label{ICZH}]{%
\includegraphics[height=5cm,width=7.5cm]{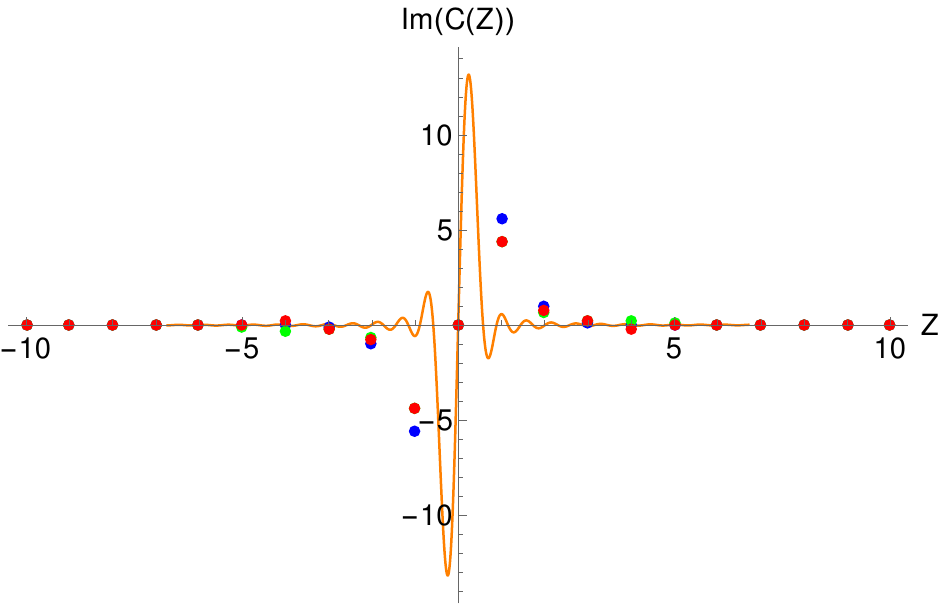}
}
\caption{Re C(Z) in (a,b)  and   Im C(Z) in (c, d) for $m/m_S=0.177$ (left panel) and $m/m_S=0.806$ (right panel) for a boost with $v=0.925$ at times $t=0.1$ (blue dots), $t=13$ (green dots), $t=25$ (red dots) using the improved lattice mass. The solid line (solid orange line) is the light front result with $v=1$ in~\eqref{DPRIMITIVEXX} using the solution to \eqref{TH1X} with 19 Jacobi polynomials. The orange line is rescaled to match the maximum of the real part at $Z=0$ (for $t=13$). For the lattice data, we fixed $N=22$ and $g=1$. The data is shown in units of $a=1$.} 
\label{DIGIT}
\end{figure*}

\begin{figure*}
\centering
\subfloat[\label{RCZ}]{%
\includegraphics[height=5cm,width=7.5cm]{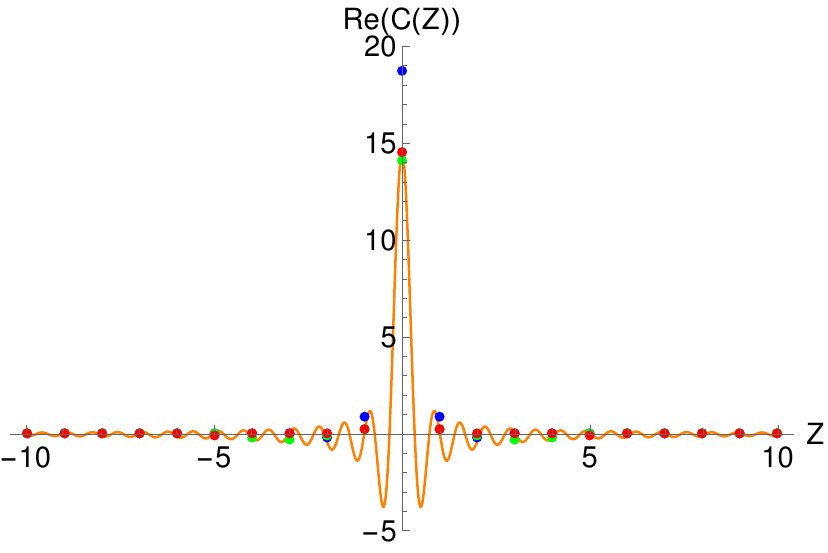}  
}\hfill
\subfloat[\label{ICZ}]{%
\includegraphics[height=5cm,width=7.5cm]{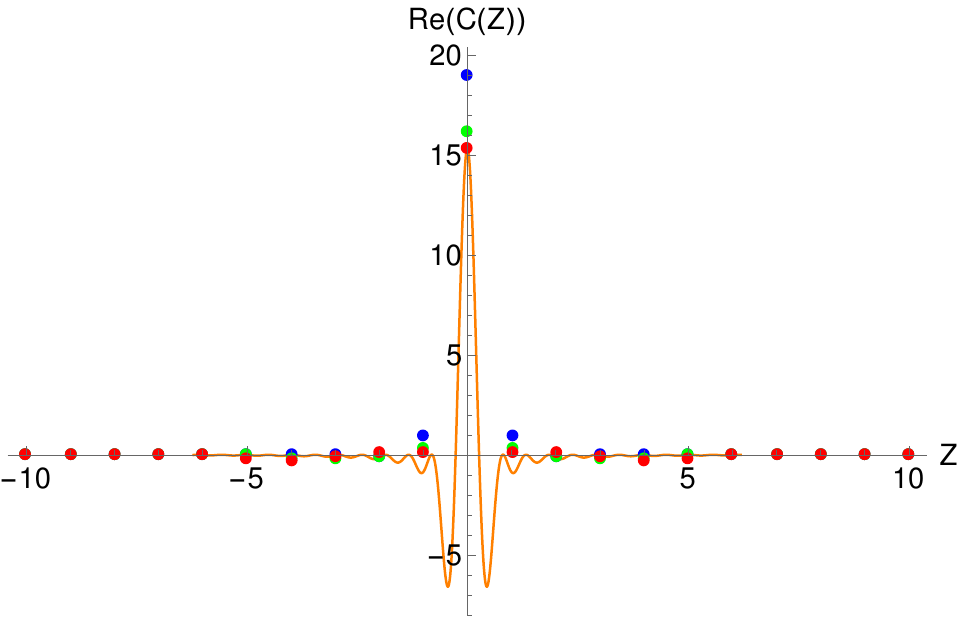}
}\hfill
\subfloat[\label{RCZH}]{%
\includegraphics[height=5cm,width=7.5cm]{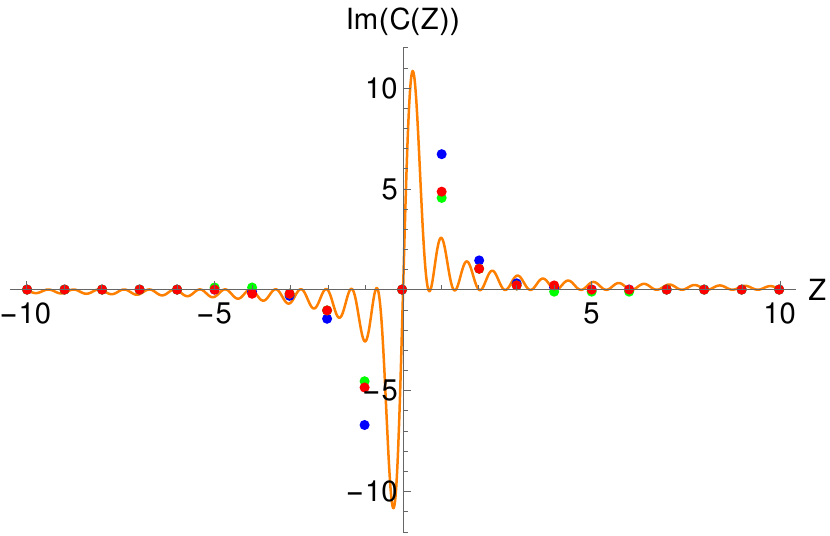}  
}\hfill
\subfloat[\label{ICZH}]{%
\includegraphics[height=5cm,width=7.5cm]{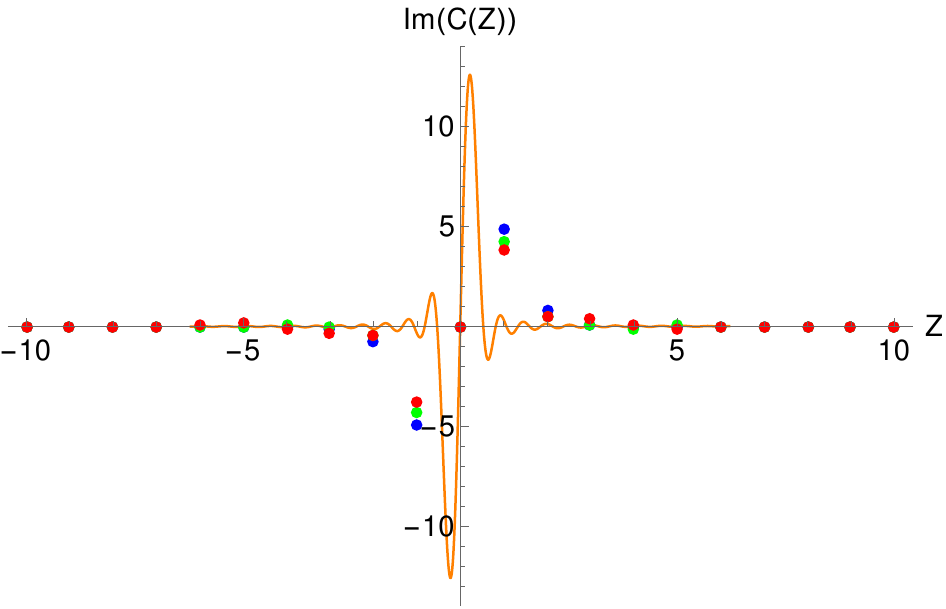}
}
\caption{Re C(Z) in a,b  and   Im C(Z) in c, d for $m/m_S=0.177$ (left panel) and $m/m_S=0.806$ (right panel) for a boost with $v=0.925$ at times $t=0.1$ (blue dots), $t=13$ (green dots), $t=25$ (red dots) using the \textit{na\"ive} lattice mass. The solid line (solid orange line) is the light front result with $v=1$ in~\eqref{DPRIMITIVEXX} using the solution to \eqref{TH1X} with 19 Jacobi polynomials. The orange line is rescaled to match the maximum of the real part at $Z=0$ (for $t=13$). For the lattice data, we fixed $N=22$ and $g=1$. The data is shown in units of $a=1$.} 
\label{fig:latticebare}
\end{figure*}

\begin{figure*}
    \centering
\includegraphics[width=0.49\linewidth]{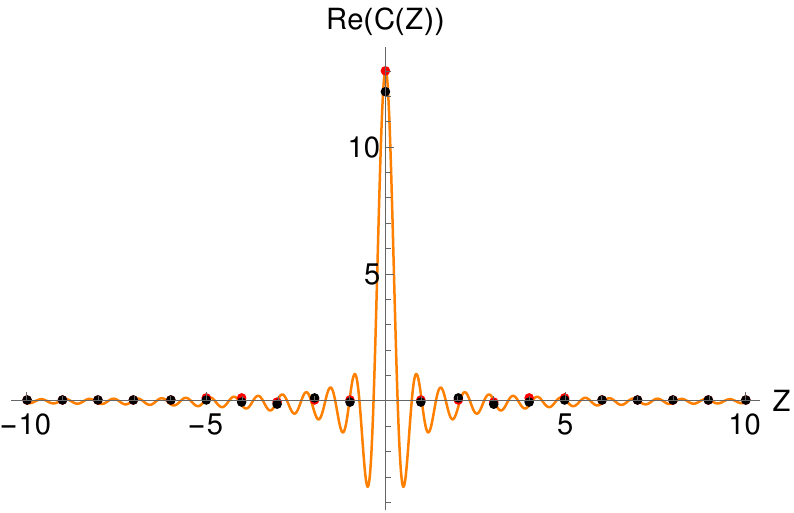} \includegraphics[width=0.49\linewidth]{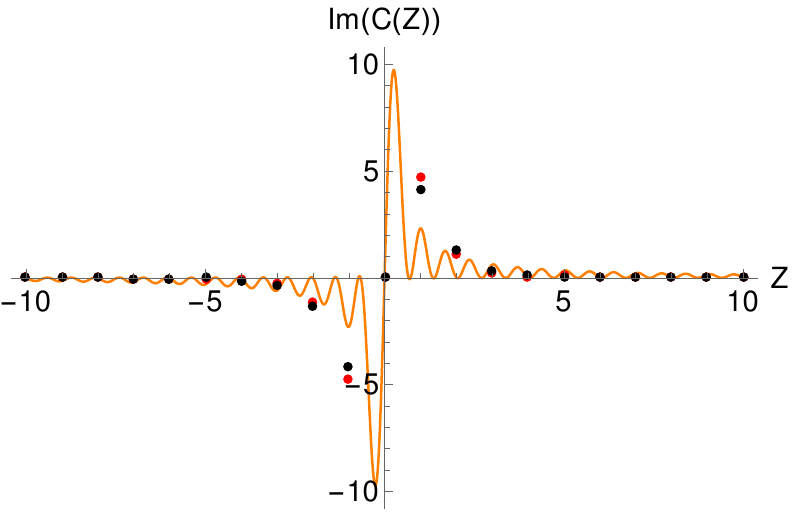}
    \caption{Comparison of the spatial qFF for the boosted (\ref{CVTDX}) (black) and unboosted form (\ref{CVTDXP}) (red). We fixed $N=22$, $t=25$, $g=1$, $v=0.925$ and $m/m_S=0.177$ using the improved lattice mass, all in units of $a=1$. }
    \label{fig:comp}
\end{figure*}

\section{Latticized Kogut-Susskind Hamiltonian}
\label{SECIV}
In this section, we explain how we discretize the Hamiltonian~\eqref{A1} on a lattice of length $L=Na$.
Using staggered fermions, the discretization of the two-dimensional massive
Schwinger model and gauge fixing is standard~\cite{Banks:1976ia}.

\subsection{Details of the lattice}
Using the Kogut-Susskind Hamiltonian formulation, the staggered fermions are mapped onto a spatial lattice by assigning the upper component of the bispinor to even sites and the lower component to odd sites, 
\bea
\psi(0,z=na)\!=\!\frac 1{\sqrt a}\!
\begin{pmatrix}
   \psi_e(n) \\
   \psi_o(n)
\end{pmatrix}
\!=\!\frac 1{\sqrt a}\!
\begin{pmatrix}
   \varphi_{n: {\rm even}} \\
   \varphi_{n+1: {\rm odd}}
\end{pmatrix}
\eea
with $0\leq n\leq N-1$. 
The chiral fermions map onto $\psi_e(n)=\varphi_n$  (even sites) and
$\psi_o(n)=\varphi_n$ (odd sites), with
\bea
\varphi_n&=&\prod_{m<n}[+iZ_m]\frac 12 (X_n-iY_n),\nonumber\\
\varphi^\dagger_n&=&
\prod_{m<n}[-iZ_m]\frac 12 (X_n+iY_n).
\eea
The boost operator maps onto
\bea
\mathbb K\rightarrow 
&&\frac 12 g^2a^2\sum_{n=0}^{N-1} nL_n^2\nonumber\\
&&+\frac 1{4}\sum_{n=0}^{N-1}\,n
\left(X_nX_{n+1}+Y_nY_{n+1}\right)\nonumber\\
&&+\frac{ma}2
\sum_{n=0}^{N-1}\,(-1)^n n\,(1+Z_n).
 \eea
Moreover, the lattice Hamiltonian reads \cite{Banks:1976ia}
\begin{align}
\begin{aligned}
\label{eq:Ham}
\mathbb{H}\to&
     \frac{1}{4a}\sum_{n=1}^{N-1}\left(X_n X_{n+1}+Y_n Y_{n+1}\right)
\\&
     +\frac{m}{2}\sum_{n=1}^N(-1)^n Z_n+\frac{a g^2}{2}\sum_{n=1}^{N-1}L^2_n,
 \end{aligned}
 \end{align}
where $L_n$ is the local electric field operator 
\begin{equation}
    L_n=\sum_{j=1}^n\frac{Z_j+(-1)^j}{2}.
\end{equation}

\subsection{Lattice qFF from Exact Diagonalization}
The discretization of the qFF in QED2 follows the same line of arguments as the one of the qPDFs in~\cite{Grieninger:2024cdl}.  The only new element is
\bea
\label{anan}
&&|\psi^\dagger \gamma_5 \psi (0,0)|^2
=\nonumber\\
&&\frac 1{a^2}\bigg|\sum_{n}
(\sigma_{n}^+\sigma^-_{n+1}-\sigma_{n+1}^+\sigma_n^-)\bigg|^2,
\eea
where $\sigma_n^\pm=\frac 12 (X_n\pm iY_n)$. With this in mind, the discretized form of the symmetric spatial qFF (\ref{CVT}) reads
\begin{widetext}
\bea
\label{CVTDX}
\mathbb C(n,v,t)=\frac 4{aF^2}\sum_{i,j=e,o}
\langle 0|\psi_i(-n)e^{i\mathbb H t} e^{i\chi(v)\mathbb K}
\,|\psi^\dagger \gamma_5\psi(0,0)|^2\,e^{-i\chi(v)\mathbb K}
e^{-i\mathbb Ht} \psi_j^\dagger(n)|0\rangle.
\eea
or equivalently
\bea
\label{CVTDXP}
\mathbb C(n,v,t)=\frac 4{aF^2}\sum_{i,j=e,o}
\langle 0|\psi_i(-n)e^{i\mathbb H t} 
\,|\psi^\dagger \gamma_5\psi(0,P(v))|^2\,
e^{-i\mathbb Ht} \psi_j^\dagger(n)|0\rangle.
\eea
\end{widetext}
with the fully boosted source
\bea
\label{Panan}
&&|\psi^\dagger \gamma_5 \psi (0,P(v))|^2
=\nonumber\\
&&\frac 1{a^2}\bigg|\sum_{n}\,e^{iP(v)na}\,
(\sigma_{n}^+\sigma^-_{n+1}-\sigma_{n+1}^+\sigma_n^-)\bigg|^2\,.
\eea

To proceed, the ground state is obtained as the lowest state of the
spin qubit Hamiltonian following from an exact diagonalization of
the Hamiltonian (\ref{eq:Ham}). Eq. (\ref{CVTDX}) is then evaluated by first applying the
double unitary evolutions, and then measuring  (\ref{anan}) at large times
$t$ for increasing rapidities $\chi$. 
In Fig.~\ref{DIGIT}, we show the numerical results for the
real (Re) and imaginary (Im) parts of the 
spatial correlator 
qFF (\ref{CVTDX}) in the strong coupling regime with 
$m/m_S=0.177$ (left panel) and the weak coupling regime with
$m/m_S=0.806$ (right panel) using the improved lattice mass~\cite{Dempsey:2022nys} as discussed in
\cite{Grieninger:2024cdl}. The results using the \textit{na\"ive} lattice mass are shown in Fig.~\ref{fig:latticebare}. 

The results for the correlators are 
at time $t=0.1$ (blue dots), $t=13$ (green dots) and $t=25$ (red dots) for a fixed boost velocity $v=0.925$ and $N=22, g=a=1$. The solid curve (solid orange line) is the light-front result with $v=1$, for the spatial correlator 
\eqref{DPRIMITIVEXX} following from the solution to \eqref{TH1X},
in the lowest Fock approximation. Since the DLY relation is an approximation to 
the FF, the comparison is only qualitative. Note that the
spatial correlator \eqref{CVTDX} is evaluated 
for large times $t\gg m_S\sim g/\sqrt{\pi}$. 
Finally, we have checked that our numerical results converge sufficiently fast above this time scale which is explicitly demonstrated in appendix~\ref{sec:convtime}.

In Fig.~\ref{fig:comp}, we compare the numerical results stemming from the boosted form (\ref{CVTDX}) and the unboosted form (\ref{CVTDXP}). The differences stem from the numerical discretization of the boost operator on this coarse lattice. In particular, we discussed the numerical errors arising from the discretized version of the boost in our previous publication~\cite{Grieninger:2024cdl}.

\section{Conclusions}	
\label{SECV}
 We have shown that the CS FFs can be obtained through equal-time and properly boosted qFFs defined
at spatial separation, with the out-going meson fragment sourced by a pertinent 
current condition at large times. This construction was explicitly implemented
in massive QED2, where the FF follows from a combination of boosts and large time
evolution. 

We have used the DLY result for the FF in QED2, to derive explicit forms for
the FFs and to analyze their behavior in the strong and weak coupling regimes. 
In the 2-Fock space approximation, the QED2 results readily extend
to QCD2 with minor changes. 

Using the Kogut-Susskind Hamiltonian on a spatial lattice with open boundary
conditions, we discretized massive QED2 using a spin formulation suitable for quantum computations. The
qFF is reduced to a Fourier transform of a spacial correlator at
fixed boost and large time, on a spatial lattice with open boundary. We have shown that for 
increasing boosts and large times the results from exact diagonalization for the qFF are
in fair agreement with the corresponding DLY correlator in the light-like limit.

The spin formulation we used, readily maps onto qubits for a full quantum computation.  More importantly, our suggested qFF can be evaluated in full QCD using 
the gauged fixed Kogut-Susskind  lattice Hamiltonian formulation with continuous time~\cite{Kogut:1974ag}. We plan to report on some of these issues next.

\begin{acknowledgments}
This work is supported by the Office of Science, U.S. Department of Energy under Contract No. DE-FG-88ER40388. The work of S.G. is supported by the U.S. Department of Energy, Office of Science, National Quantum Information Science Research Centers, Co-design Center for Quantum Advantage (C2QA) under Contract No.DE-SC0012704. S.G. is also in part supported by a Feodor Lynen Research fellowship of the Alexander von Humboldt foundation. This research is also supported in part within the framework of the Quark-Gluon Tomography (QGT) Topical Collaboration, under Contract No. DE-SC0023646.
\end{acknowledgments}

\appendix
\section{$\eta$ DA and PDF}
\label{DAPDF}
In the 2-particle Fock-space
approximation, the light front wave functions
DAs $\varphi_n(x)$ for QED2 follow from~\cite{Bergknoff:1976xr}
\begin{widetext}
\bea
\label{TH1X}
\mu^2\varphi(x)=\int_0^1dy\,\varphi(y)+\frac {\alpha}{x\bar x}\varphi(x)-\,{\rm PP}\int_0^1dy\,\frac{\varphi(y)-\varphi(x)}{(x-y)^2}
\eea
\end{widetext}
with $x P$  being the  momentum fraction of the partons. Here PP is short for the principal part, and the  masses are rescaled
$$\mu^2=M^2/m_S^2,\qquad\alpha=m^2/m_S^2.$$ 
The longitudinal kinetic contribution (second term on the RHS)  is singular at $x=\pm 1$, forcing the light front wave function to vanish 
at the edges $\varphi_n(\pm 1)=0$.

In the massless limit with $m\rightarrow 0$, the 
spectrum is that of a single massive boson
\bea
\label{PHIM0}
\varphi_\eta(x )=\varphi_0(x)\rightarrow \theta(x\bar x),\qquad M_0^2\rightarrow m_S^2.
\eea
 For small $m$,  the lowest solution vanishes at the end points as powers of $m$, $\varphi_n(x)\sim (x\bar x)^\beta$ 
with $\beta$ fixed by~\cite{tHooft:1974pnl}
\bea
\frac{m^2}{m_S^2}=1-\pi\beta{\rm cotan}(\pi\beta).
\eea
In the massive limit with $m\gg m_S$, the solution is peaked around $x=\frac 12$, with $M\approx 2m$. The lowest state parton distribution function is
of the form
\bea
\label{QM0}
q_\eta(x )=|\varphi_\eta (x )|^2.
\eea

The general solution to (\ref{TH1X}) can be obtained though numerical diagonalization, using the basis expansion~\cite{Mo:1992sv} 
\bea
\label{BASISEXP}
\varphi(x)=\sum_n \bar c_n \bar f_n(x)
\eea
with Jacobi polynomials
\bea
\label{FBARN}
\bar f_n(x)=\bar C_n(x\bar x)^\beta P_n^{(2\beta, 2\beta)}(x-\bar x)
\eea
normalized through
$$
\bar C_n=\bigg(n! (1+2n+4\beta)
\frac{\Gamma(1+n+4\beta)}
{\Gamma^2(1+n+2\beta)}\bigg)^{\frac 12}.
$$
The matrix form of (\ref{TH1X}) is
\bea
\label{MAT1X}
(\bar{\mathbb A}_{mn}+\bar{\mathbb B}_{mn}+\bar{\mathbb  C}_{mn})\,\bar c_n=\mu^2\,\bar c_m
\eea
with the matrix entries
\bea
\label{MATENT}
\bar{\mathbb A}_{mn}&=& 
\int_{0}^1\,dxdy\,\bar f_m(x)\bar f_n(y)
\nonumber\\
\bar{\mathbb B}_{mn}&=& \alpha\int_{0}^1\,dx\,\frac{\bar f_m(x)\bar f_n(x)}{x\bar x}\nonumber\\
\bar{\mathbb C}_{mn}&=& 
-{\rm PP}\int_{0}^1\,dxdy\,
\frac{\bar f_m(x)(\bar f_n(y)-\bar f_n(x))}{(x-y)^2}.\nonumber\\
\eea
The procedure for the diagonalization was recently detailed in~\cite{Grieninger:2024cdl} (and references therein).

\begin{figure}
    \centering    \includegraphics[height=5cm,width=0.99\linewidth]{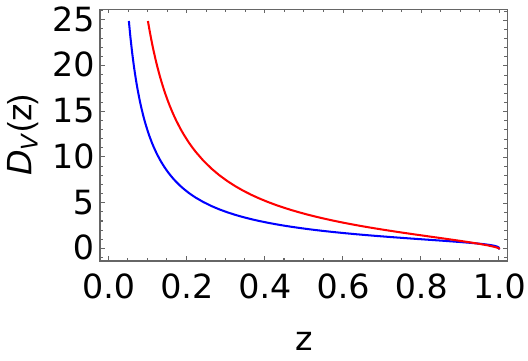}
    \caption{Fragmentation function for light quarks from QCD2, following from the leading planar contribution in (\ref{EIN1}) from~\cite{Einhorn:1976ev}.}
   \label{DLYLIGHT2}
\end{figure}

\section{DLY in  QCD2}
Given the similarity between (\ref{TH1X}) 
in the 2-Fock space approximation, and
the 't Hooft equation for QCD2 in the large $N_c$ limit on the light front (suppressed vacuum contributions), we conclude that 
the DLY fragmentation of a quark 
$Q_f\rightarrow Q_f+n$ in QCD2 gives
\bea
\label{D1FINITEQCD2}
&&d_{DLY}^n(z,1)\rightarrow \nonumber\\
&&\frac{1}{N_c}\frac{\bar z^2}
{z(\bar z\mu^2+z^2\bar\alpha)^2}
\bigg(\int_0^1dx \frac{\varphi_n(x)}{(x-1/z)^2}\bigg)^2
\eea
and $m_S^2\rightarrow \lambda/2\pi$
with fixed 't Hooft coupling $\lambda=g^2N_c$. Recall that in QCD2  there is no $U(1)$ anomaly, so the $f$-contribution drops out from the squared bracket in (\ref{D1FINITEQCD2}). 
This result is in agreement with the one recently derived in~\cite{Jia:2023kiq}, using the $1/N_c$ counting in light front perturbation theory.

In
the massless limit, the 't Hooft equation admits a massless mode
with $\varphi_0(x)=\theta(x\bar x)$ and $\mu=0$ (would-be Goldstone mode), for which (\ref{D1FINITEQCD2}) simplifies
\bea
\label{D1FINITEQCD20}
d_{DLY}^0(z,1)\rightarrow \frac 1{N_cz}
\eea
and normalizes to $1/N_c$ for each quark color.
In the massive but strong coupling case, the massless mode turns massive with $\varphi_0(x)\approx (x\bar x)^\beta$ and $\mu^2\approx m/m_S$. Eq. (\ref{D1FINITEQCD2}) is apparently singular, but reverting  to the original DLY form
\bea
\label{D1FINITEQCD2X}
d_{DLY}^n(z,1)\rightarrow \frac 1{N_cz}\bigg|\varphi_n\bigg(\frac 1z, 1\bigg)\bigg|^2
\eea
yields a similar result to QED2 under the substitution $\varphi_0(x)\approx (x\bar x)^\beta$ for light quark masses. In the heavy mass limit, the DLY fragmentation function for QCD2 is in total agreement with QED2, once the momentum sum rule is enforced.

\begin{figure*}
    \centering
\includegraphics[width=0.32\linewidth]{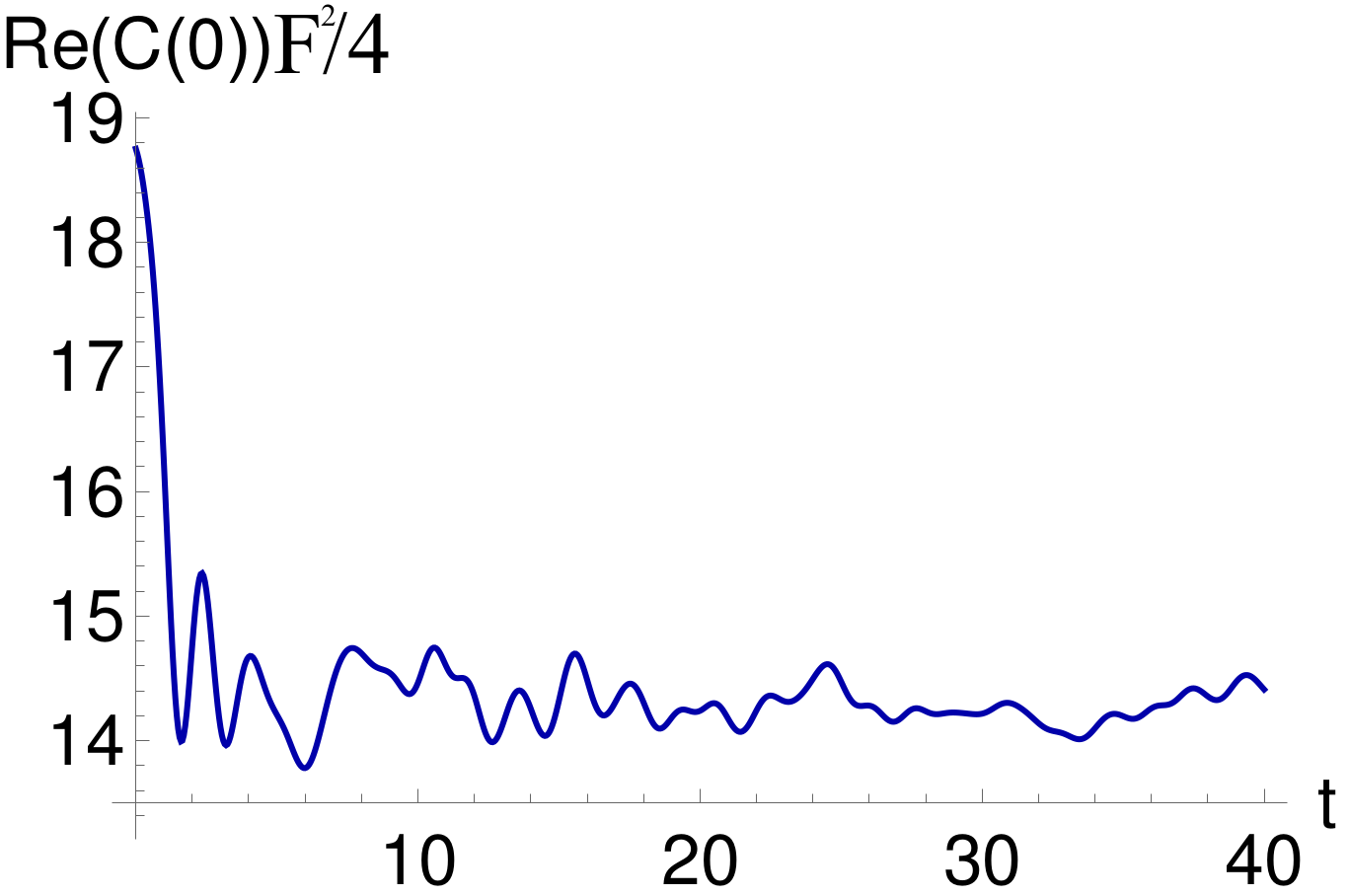} \includegraphics[width=0.32\linewidth]{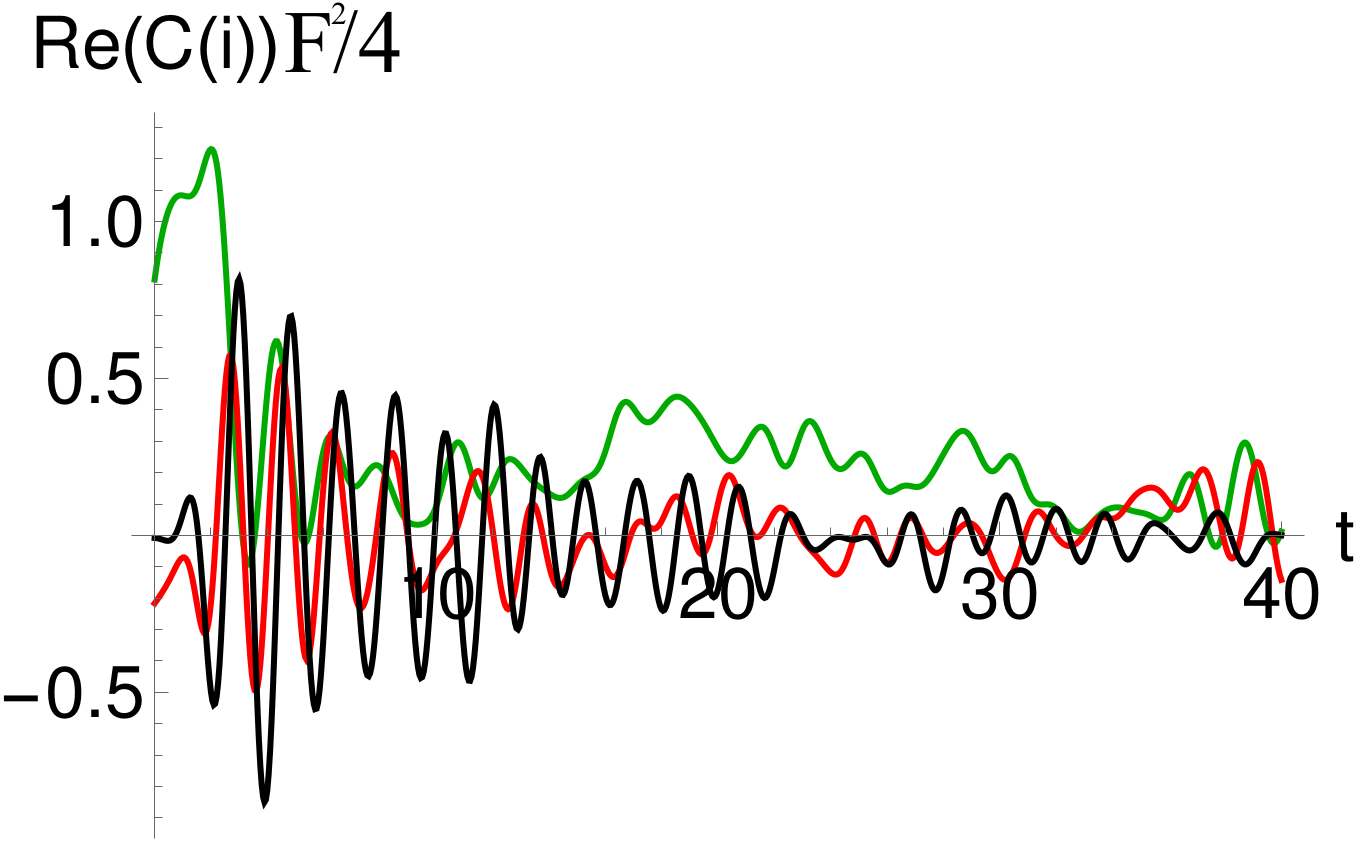}
\includegraphics[width=0.32\linewidth]{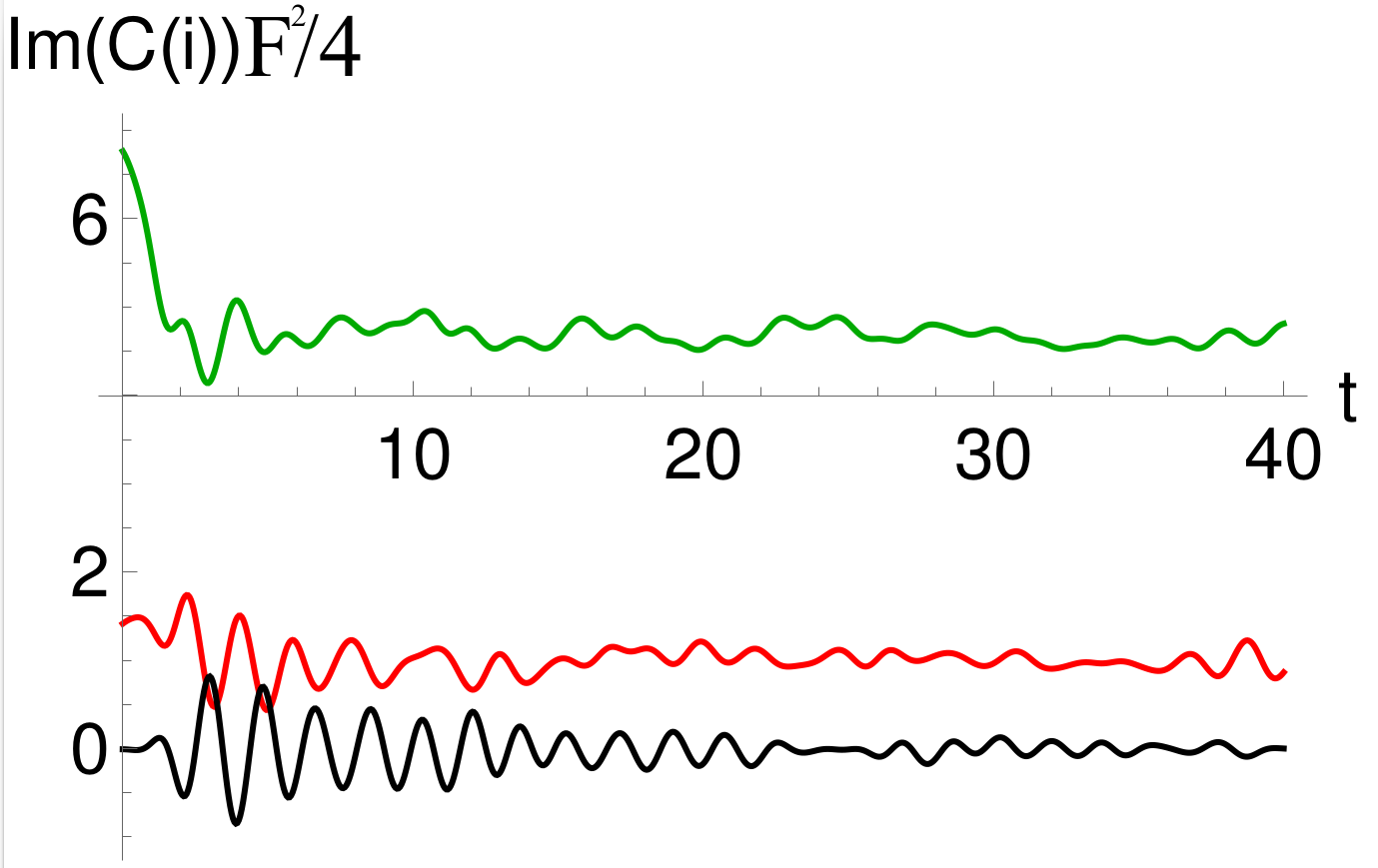}
    \caption{\textit{Left:} Re(C(0)), \textit{Middle:} Re(C(1)) (green), Re(C(2)) (red), Re(C(3)) (black), \textit{Right:} Im(C(1)) (green), Im(C(2)) (red), Im(C(3)) (black), as a function of time. We fixed N=18, $g=1$ and $m/m_S=0.806$ using the improved lattice mass; all plots are in units of $a=1$. }
    \label{fig:apptime}
\end{figure*}

The result (\ref{D1FINITEQCD2}) is of order $1/N_c$ for each quark color and flavor
and sums up to $N_c^0$ at variance with expectations  from  duality where it should reduce to the famed R-ratio of order $N_c$. Indeed, in~\cite{Einhorn:1976ev} it was argued that the DLY relation in QCD2 does not yield the expected R-ratio scaling in the 't Hooft limit. In this limit, the annihilation $e^+e^-\rightarrow X$ in terms of planar diagrams proceeds solely through meson decays. For a meson decay  to a pair of mesons $V\rightarrow V+n$, the result is~\cite{Einhorn:1976ev}
\begin{widetext}
\bea
\label{EIN1}
\frac{d{\rm ln}\sigma_n}{dz}=
(d_V^n(z)+d_{\bar V}^n(z))=
\frac{C_n}z\bigg(
\bigg|\frac 1z \int_z^1dy\int_0^1dx\frac{\varphi_n(x)}{(x-y/z)^2}\bigg|^2
+
\bigg|\frac 1z \int_0^{\bar z} dy\int_0^1 dx\frac{\varphi_n(x)}{(x-(y-\bar z)/z)^2}\bigg|^2
\bigg).\nonumber\\
\eea
\end{widetext}
with $d_V^n$ and $d_{\bar V}^n$ identified with the first and second contributions, respectively.
It is of order $N_c^0$ as well for each meson state lying on the radial
Regge trajectory (see below). 
Here $C_n$ is fixed by the momentum sum rule
\bea
\label{EH2}
\int_0^1 dz\,\frac{d{\rm ln}\sigma_n}{d{\rm ln}z}=1.
\eea
For small quark masses, we can substitute the approximation $\varphi(x)\rightarrow (x\bar x)^\beta$ for the $n=0$ ground state, and carry the integration in (\ref{EIN1}) with the result
\bea
\label{EH3}
d^0_V(z)={C}\frac {\bar z}z\Gamma[2\beta]\Gamma[1+2\beta]
\,{}_2\tilde F_1[1, 1 + 2 \beta, 1 + 4 \beta, z]\nonumber\\
\eea
where $\tilde F$ is the regularized confluent hypergeometric function, and with the normalization  fixed through
 \begin{widetext}
 \bea
 \label{EH4}
 4C\beta\Gamma[2\beta]\Gamma[1+2\beta]
 \frac{(1 + 1/(1 - 4\beta) + \psi(2 \beta )- \psi(4 \beta -1))}
 {(-1 + 2\beta)\Gamma[1 + 4\beta]}=1,
\eea
 \end{widetext}
where $\psi(z)$ is the digamma function. Eq. (\ref{EIN1}) indicates that the differential
cross section for the inclusive process is also 
characterized by $d_V^n$ FFs, as Feynman suggested
in the parton model proposal. However, these FFs follow solely from the planar decays of  
mesons in the large $N_c$ limit, and not from crossing the leading order handbag diagram
as per weak coupling and factorization in the light front analysis, hence the difference between $d_V^n(z)$ and $d_{DLY}(z)$ which follows from the crossing argument.

Remarkably, \eqref{EIN1}-\eqref{EH4} exhibit scaling with the end point behavior $d_V^n(z\sim 1)\sim \bar z^{2\beta}$  and $d_V^n(z\sim 0)\sim 1/z$, as per the QCD2 DLY result (\ref{D1FINITEQCD2}), and also in agreement with QED2. However, the overall coefficients
multiplying the scaling laws are different.
In Fig.~\ref{DLYLIGHT2} we show the light quark fragmentation functions in QCD2 following from \eqref{EIN1}-\eqref{EH2}, for $\beta=0$ (solid blue line) and $\beta=0.2$ (solid red line), which is to be compared to Fig.~\ref{DLYLIGHT} for QED2. The behaviors are comparable but not identical.
 
 Additional planar contributions from splittings of the fragmented mesons to additional mesons contribute to the same order. They correspond to a hadronic out-in cascade of mesons
in the large number of colors limit~\cite{Einhorn:1976ev}.

For the meson pair decay, the total inclusive cross section in QCD2 in the 't Hooft limit  yields the expected R-ratio scaling, when the sum over all intermediate meson resonances
with width $1/N_c$ is carried out for fixed $\sqrt s$~\cite{Einhorn:1976ev}, 
\bea
\sigma_{e^+e^-}=\sum_{n,f}e_f^2\sigma_n\rightarrow R=
\sum_f N_ce_f^2
\eea
The R-ratio is dominated by the short distance UV physics. The same observation was made for the F2 structure function~\cite{Einhorn:1976ev}.
The scaling laws in QCD2 are recovered  from free field theory,
when the 't Hooft limit is carried carefully. In QCD4 they follow from 
asymptotic freedom at weak coupling.

\section{Convergence of $C(Z)$ with time}\label{sec:convtime}

In order to get an understanding of how much the qFF at a fixed lattice site varies with asymptotic time, we show the time evolution of the four inner most points ($Z=0,1,2,3$). From Fig.~\ref{fig:apptime} it is clear that the maximum at $Z=0$ rapidly converges to a constant with small oscillations.

\bibliography{qed2parton}
\end{document}